%% file: main.tex
\documentclass[twocolumn, secnumarabic,  floatfix, nofootinbib, nobalancelastpage]{revtex4-2}
\pdfoutput=1

\def\TitleOfPaper{Machine learning and density functional theory}
\def\preprintlinklocation{http://dft.uci.edu + PAPER REF}

\input Burke_template

\input KieronsMacros

\graphicspath{{figures/}}

\definecolor{Green}{rgb}{0.016,0.627,0}

\usepackage[encapsulated]{CJK}

\begin{document}
\begin{CJK*}{UTF8}{gbsn}
\sf
\coloredtitle{\TitleOfPaper}


\coloredauthor{Ryan Pederson}
\affiliation{Department of Physics and Astronomy, University of California, Irvine, CA 92697, USA}

\coloredauthor{Bhupalee Kalita}
\affiliation{Department of Chemistry, University of California, Irvine, CA 92697, USA}

\coloredauthor{Kieron Burke}
\email{kieron@uci.edu}
\affiliation{Department of Chemistry, University of California, Irvine, CA 92697, USA}
\affiliation{Department of Physics and Astronomy, University of California, Irvine, CA 92697, USA}

\date{\today}
\begin{abstract}
Over the past decade machine learning has made significant advances in approximating density functionals, but whether this signals the end of human-designed functionals remains to be seen. Ryan Pederson, Bhupalee Kalita and Kieron Burke discuss the rise of machine learning for functional design.
\end{abstract}
\maketitle
\end{CJK*}



In 2016, DeepMind's AlphaGo made history as the first computer program to defeat a professional human player in the board game Go. Aside from excelling at board games with human-designed rules, machine learning (ML) has seen impressive applications in scientific research~\cite{douglas2022machine} where the rules are instead governed by nature, such as protein folding and, more recently, density functional theory (DFT). But can the ML approach to DFT approximations significantly outperform human-designed ones and solve long-standing challenges?

\sec{DFT everywhere}
In physical sciences, DFT is often the go-to computational method for solving electronic structure problems. DFT provides fully quantum solutions at a fraction of the cost of solving the Schrödinger equation directly, by mapping the coupled many-body problem to a single-particle problem. The electronic energy is considered as a functional (a function of a function) of the electron (probability) density, with only a small portion, the exchange-correlation energy, being approximated. 

It is staggering to see just how important DFT calculations have become. Each year, tens of thousands of papers report useful predictions from DFT calculations, and today about one third of the National Energy Research Scientific Computing Center supercomputing resources use DFT ~\cite{nersc10} . John Perdew, who developed many of the formulas in current use, is one of the most cited physicists of all time.

\sec{The GIGO principle in DFT}

The GIGO principle is an old adage in computing standing for: garbage in, garbage out. In DFT this means that a calculation is only as good as the approximate functional used. Humans have worked at this for almost a century, and nowadays hundreds of different approximations are in use. Some build in well-studied limits, such as the uniform electron gas, and satisfy many known physical constraints of the exact functional, while others are tuned and fitted to reference datasets. Regardless, general failures have been identified over the years. A decade-old review~\cite{cohen2012challenges} focused on the struggle to describe strongly correlated systems. This most grievous failure can be understood from the perspective of fractional charges (systems with noninteger total charge) and fractional spins (systems with noninteger spin magnetization). The exact energy is a linear interpolation of the energy of the adjacent integer systems, but approximations miss this, producing embarrassingly large systematic errors in strongly correlated systems as simple as stretched H$_2$. Overcoming such fundamental DFT challenges is essential to expanding its applicability and reliability in condensed matter physics.

\sec{Machine learning DFT}

A proof of principle for ML-DFT appeared 10 years ago. For a simple problem, the kinetic energy of non-interacting fermions in a 1D box, a ML method (kernel ridge regression) could be used to find an approximation of the functional by training on examples from accurate numerical calculations~\cite{snyder2012finding}. The resulting functional was far more accurate than anything ever designed by humans, but only useful for simple model systems such as those it was trained on. The associated learning efficiency was also low, as hundreds of training examples were needed to reach high accuracy for a rather compact chemical space. 

Later, the density was machine-learned directly from the external potential~\cite{BVLM17}. This demonstrated the practical usefulness of ML in DFT through realistic examples, such as proton transfer in a ML molecular dynamics simulation of malonaldehyde. However, unlike traditional DFT approximations, such ML models rarely generalize across elements.

Since then, there have been many attempts to bring the promise of ML to practical, generalizable functional construction. These efforts can be divided into two categories:  those starting from traditional approximate forms suggested by humans (which are biased toward local and semi-local approximations) and those that use the entire density (that is, a non-local approximation) in some hard-to-understand way. Such non-local functionals can have poor generalizability, as the input feature space becomes vastly more complicated than local and semi-local forms which depend only on the density and its gradient at each point.

As described in Ref.~\cite{nagai2020completing} a neural network (NN) functional was trained on accurate densities as well as energies of just three molecules, producing  semi-local ML approximations that worked as well as human-designed functionals for 150 test molecules, generalizing very well. A similar approach was used in Ref.~\cite{li2021kohn} but non-local forms based on convolution NNs were also used to learn an entire dissociation curve within chemical accuracy, including the strongly correlated region, with only two training examples. The model also generalized well for other new (but similar) strongly correlated molecules that were not encountered in training. During training, an end-to-end differentiable DFT code (where all components are differentiable) was used to obtain gradient information by backpropagation through the entire self-consistent calculation. Such robust gradient-based training results in impressive generalization of functional approximations.

But the most recent exciting development comes once again from DeepMind~\cite{kirkpatrick2021pushing}.  A bevy of 17 researchers, using vast computational resources, revived an old human-designed suggestion, a local hybrid functional~\cite{cruz1998exchange}, that had been difficult to control. Their new NN-based functional, DM21, was trained by evaluating the energy non-self-consistently using approximate densities. The regression loss consisted of an energy loss plus an explicit gradient regularization term, thereby making this training approach substantially cheaper than Ref.~\cite{li2021kohn}. DM21 was trained on thousands of molecular systems, orders of magnitude more than previous ML training sets, and outperforms most other hybrid functionals on standard molecular benchmarks with impressive generalization. This ML functional can be used for main-group chemistry calculations, like most human-designed functionals. By including training on simple systems with fractional charges and spins, DM21 appears to perform significantly better than earlier approaches for strongly correlated systems. For instance, DM21 correctly dissociates systems such as H$_2$, H$_2^+$, and N$_2$, meeting the long-standing DFT challenge of strong correlation in molecular systems.\\

\sec{Will DFT go the way of Go?}
Researchers all over the world are currently trying out DM21, testing many different aspects to see if it lives up to its promise.  The world of DFT applications is far too vast for DM21 developers to run even a fraction of useful tests in their original paper.  Many promising approximations run into unexpected difficulties when tried in practice. The community will examine computational cost, accuracy, and transferability when testing DM21.

A basic issue is whether DeepMind's approach can also work for materials. DM21 was trained and tested only on molecules, where wave function-based quantum chemistry provides high accuracy benchmark data. If such data were available for materials, for example, via computationally-expensive Quantum Monte Carlo simulations, a similar performance might be expected. However, the real aim of DFT is to find a single functional that works for both molecules and materials simultaneously, so that you can calculate everything in between, such as surfaces, clusters, and defects. Harsh experience suggests that accuracy for molecules degrades when good performance for solids is also required ~\cite{perdew2021artificial}.

But, in the grand scheme of things, does this development signal the beginning of the end for human-designed functionals?  Just as the world’s best Go player cannot compete with AlphaGo, can human insight and ingenuity long survive against huge data benchmarks, teams of coders, and almost unlimited computational resources? Or will human insight and ingenuity always be needed to choose the forms that machines learn from? The next decade will likely answer this question.

\bibliographystyle{apsrev4-2}
\bibliography{Master}

\sec{Acknowledgments}
Work supported by DOE DE-SC0008696 (R. P.) and NSF CHE-2154371 (B. K., K. B.).

\sec{Competing interests}
The authors declare no competing interests.


\end{document}

%% file: Burke_template.tex
\usepackage{amsmath}
\usepackage{physics}
\usepackage{amssymb}
\usepackage{hyperref}
\usepackage{float}
\usepackage[dvipsnames]{xcolor}

\usepackage[caption=false]{subfig}
\usepackage{tabularx}

\definecolor{TITLECOL}{rgb}{0.05,0.25,0.85}
\definecolor{CONTENTSCOL}{rgb}{0.1,0.2,0.7}
\definecolor{URLCOL}{rgb}{0,0.52,0.83}
\definecolor{LINKCOL}{rgb}{0.05,0.5,0}
\definecolor{CITECOL}{rgb}{0.25,0,0.48}
\definecolor{SECOL}{rgb}{0.07,0.31,0.80}
\definecolor{SSECOL}{rgb}{0.26,0.19,0.75}

\newcommand{\coloredtitle}[1]{\title{\textcolor{TITLECOL}{#1}}}
\newcommand{\coloredauthor}[1]{\author{\textcolor{CITECOL}{#1}}} 
\renewcommand{\sec}[1]{\section{\textcolor{SECOL}{#1}}}

\usepackage{graphicx}

\def\PublicationsLink{http://dft.uci.edu/publications.php}
\def\preprintlink{ \href{\preprintlinklocation}{\TitleOfPaper} }
\def\preprinttext{~}

\usepackage{fancyhdr}
\makeatletter
\fancypagestyle{titlepage}
{	\lhead{\textsc{\preprinttext\ ~ \href{\PublicationsLink}{~}}}
	\chead{}
	\rhead{\preprintlink}
	\lfoot{}}
\makeatother
\def\preprintlink{ 
	\href{\preprintlinklocation}
        {
~}
	}
	
\usepackage[normalem]{ulem}
\usepackage{xcolor}

\newcommand{\stkout}[1]{\ifmmode\text{\sout{\ensuremath{#1}}}\else\sout{#1}\fi}

\usepackage{verbatim}

%% file: KieronsMacros.tex
\def\sss{\scriptscriptstyle\rm}

\def\1var{(\bx_1...\bx\N)}

\def\bx{{x}}

\def\N{_{\sss N}}

\def\sph_int{ {\int d^3 r}}
